\documentclass[12pt]{iopart}
\usepackage{graphicx} 
\begin{document}

\def\la{\langle}
\def\ra{\rangle}
\def\om{\omega}
\def\Om{\Omega}
\def\vep{\varepsilon}
\def\wh{\widehat}
\def\P0{\wh{\cal P}_0}
\def\dt{\delta t}
\newcommand{\beq}{\begin{equation}}
\newcommand{\eeq}{\end{equation}}
\newcommand{\beqa}{\begin{eqnarray}}
\newcommand{\eeqa}{\end{eqnarray}}
\newcommand{\intf}{\int_{-\infty}^\infty}
\newcommand{\into}{\int_0^\infty}
\title[Optimal atomic detection]{Optimal atomic detection
by control of detuning and spatial dependence of laser intensity}
\author{B. Navarro*\dag, I. L. Egusquiza*,
J. G. Muga\dag\, and 
G. C. Hegerfeldt\ddag}
\address{* Fisika Teorikoaren Saila, Euskal Herriko Unibertsitatea,
644 P.K., 48080 Bilbao, Spain}
\address{\dag Departamento de Qu\'\i{}mica-F\'\i{}sica, Universidad del
Pa\'\i s Vasco, Apdo. 644, 48080 Bilbao, Spain}
\address{\ddag Institut f\"ur Theoretische Physik, Universit\"at
G\"ottingen, Bunsenstr. 9, 37073 G\"ottingen, Germany}

\begin{abstract}
Atomic detection by fluorescence may fail because of 
reflection from the laser or transmission without excitation. 
The detection probability for a given velocity range may be improved   
by controlling the detuning and the spatial dependence of 
the laser intensity.
A simple optimization method 
is discussed and exemplified. 
\end{abstract}
\pacs{03.65.-w, 42.50-p, 32.80-t}
\maketitle
\section{Introduction}

One of the standard ways to measure the time of flight, 
or simply the presence of an atom, consists of illuminating it with
a laser 
and detect the  
induced fluorescence. 
There are many different experimental settings depending 
on the incident atomic velocities and spatial span of the atomic
cloud or beam.
In particular,    
the light of the probe laser 
may be spread as a broad sheet perpendicular to the atomic motion
or be focused 
onto a diameter of a few microns; it may also be continuous or pulsed.
\begin{figure}
\begin{center}
{\includegraphics[width=11cm]{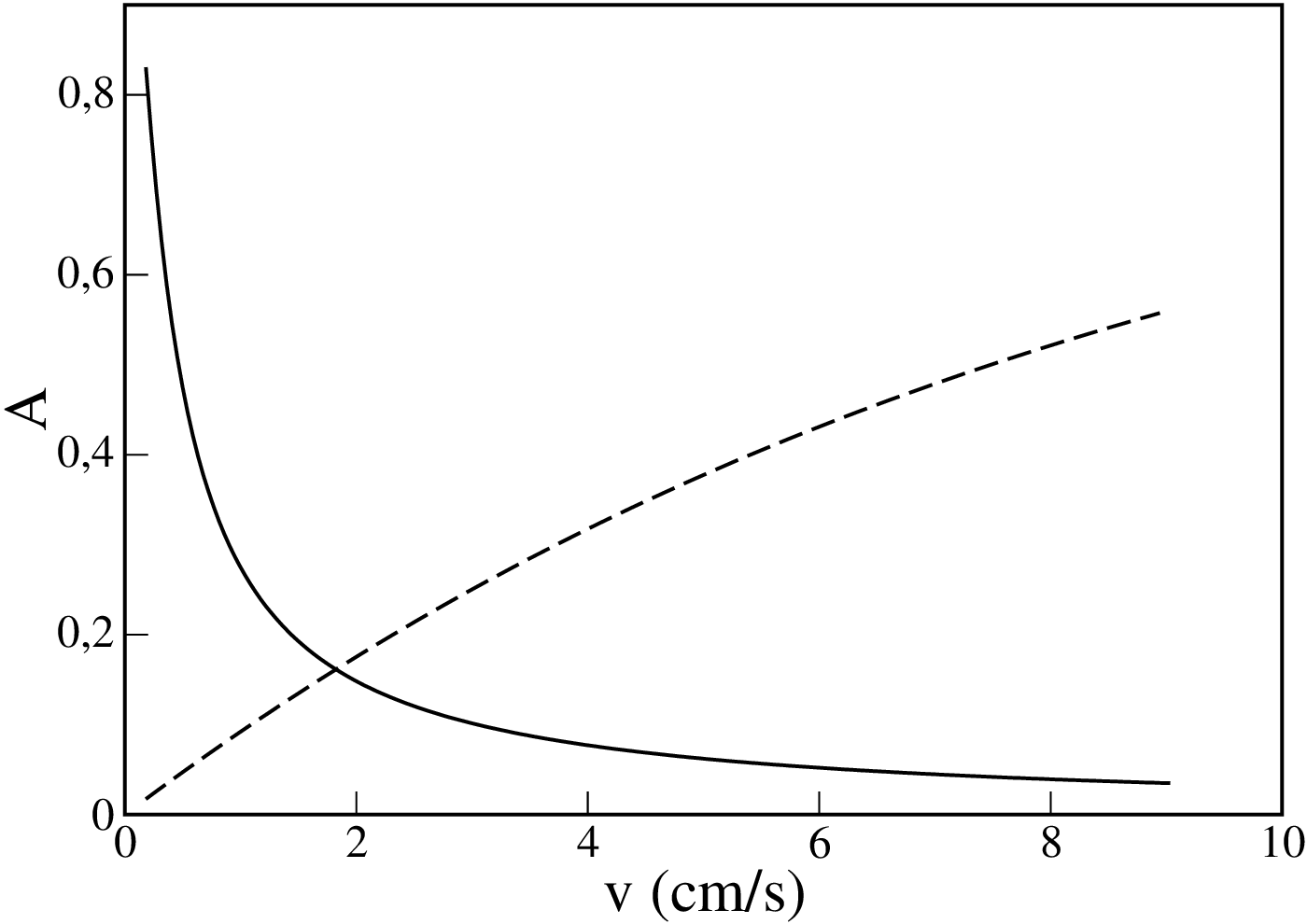}}
\caption{Detection probability versus velocity $v$ 
for laser beam width $L=10\,\mu$m.  
(This and other figures are obtained for the 
transition at 852 nm of Cs atoms, with
$\gamma=33.3\times 10^{6}$ s$^{-1}$.)
$\Omega=0.1033\times 10^6$
s$^{-1}$ (solid line); 
$\Omega=5\gamma$ (dashed line); in both cases $\Delta=0$. 
}
\label{f1}
\end{center}
\end{figure}
These detection schemes may fail however at low (ultracold)
atomic velocities
because of atomic reflection from a strong laser
field;    
the atoms may also traverse 
the finite laser-illuminated region without emitting 
any photon even at moderate velocities if the laser intensity 
is weak.  

Figure 1 illustrates these two phenomena. It represents
the detection probability versus (ultracold)
atomic velocity for a given laser-beam
width $L$ and two different laser intensities.
This and other figures below are for Cs atoms, and the transition 
$S_{1/2},\,F=4\to P_{3/2},\,F=5$, with Einstein coefficient 
$\gamma=33.3\times10^{-6}$ s$^{-1}$.
The minimum velocity considered is $0.2$ cm/s, slightly below the recoil 
velocity $\sim 0.35$ cm/s. 
In each case the laser intensity has been assumed to be constant 
in the illuminated region for simplicity. Note the significant 
atomic reflection   
for strong driving and low velocities. By contrast the weak laser
detects the very slow atoms but fails for the faster ones. 
No intermediate intensity between the two cases depicted can 
avoid the two problems 
simultaneously in an arbitrarily broad velocity range.   
An ideal detection should be able to avoid both 
effects at least for the velocity spread of the atomic clouds
of interest.  
The objective of this paper is to show that an 
appropriate adjustment of detuning and of the position dependence of the  
laser intensity   
may indeed achieve this goal.  

In the next section we shall present the theory to model the atom 
detection, section 
III describes the 
optimization of laser intensity and detuning, and in section IV 
we provide a numerical example.

\section{Basic theory}

The fundamental theory for the modeled experiment is  
described in Refs. \cite{DEHM02}, \cite{NEMH03} and \cite{DEHM03}.   
A two-level atom with transition frequency $\omega$
impinges along the $x$ direction 
on a perpendicular 
continuous laser beam of frequency $\omega_L$
directed along the $y$ direction.
In the so called
quantum jump approach  \cite{Hegerfeldt92} the continuous
measurement of the fluorescence 
photons is simulated by a periodic projection onto no-photon or 
one-photon subspaces every $\Delta t$, a time interval large enough
to avoid the Zeno effect, but smaller than any other characteristic time.
The amplitude for the
wavepacket of undetected atoms in the interaction picture
for the internal Hamiltonian 
obeys, 
in a time scale coarser than $\Delta t$, and using the
rotating wave and dipole 
approximations, an effective Schr\"odinger equation
governed by the complex ``conditional''  Hamiltonian
(the hat is used to distinguish momentum and position operators from 
the corresponding c-numbers)
\beqa\nonumber
H_{{\rm c},3D} &=& {\bf{\hat p}}^2/2m + \frac{\hbar}{2}
\Omega({\hat x})\, 
\left\{ |2\rangle\langle1| e^{i k_L \hat{y}}+
{\rm h.c.}\right\}
\\
\label{2.4}
&-&\frac{\hbar}{2}(2\Delta+i\gamma)|2\rangle\langle2|,
\eeqa
where $\gamma$ is the Einstein 
coefficient of the excited level (level 2),
i.e. its decay rate or inverse life time,
$\Omega(x)$ is the position dependent Rabi frequency
(assumed to be real), $\Delta=\omega_L-\omega$ is the detuning, 
and ${\bf{\hat p}}$ is the 
momentum operator in three dimensions (3D). The  factor
$e^{i k_L \hat{y}}$
takes into account  the spatial dependence of the laser coupling.
A one dimensional model is obtained by 
assuming that the atomic
wave packet is centered at $y=0$ and satisfies $k_L\Delta y\ll 1$
before the first photon emission, 
so that the exponentials can be dropped 
and a one dimensional kinetic term  suffices \cite{NEMH03},   
\begin{equation}\label{2.7}
H_{\rm c} = \hat{p}^2/2m 
+ \frac{\hbar}{2}\,
\left({0\atop \Omega({\hat{x}})}\;\;\;\;\;{\Omega(\hat{x})
\atop -i\gamma-2\Delta}\right), 
\end{equation}
where $\hat{p}$ is the momentum operator conjugate to $\hat{x}$,    
the 
ground state $|1\rangle$ is in vector-component notation ${1 \choose 0}$,
and the excited state $|2\rangle$ is ${0 \choose 1}$.  
%
%
The probability, $N_t$, of no photon detection 
from $t_0$, the instant when the packet is prepared far from the laser 
and with positive momenta, up to time $t$,
is given by \cite{Hegerfeldt92}
\begin{equation} \label{2.5}
N_t = || e^{-i H_{\rm c}(t-t_0)/\hbar} |\psi (t_0)\rangle||^2,
\end{equation}
where $|\psi(t_0)\ra$ is the (two-component) wave vector at $t_0$,  
and the probability density, $\Pi (t)$, for the first photon detection
by
\begin{equation} \label{2.6}
\Pi (t) = - \frac{dN_t}{dt} = \gamma P_2,
\end{equation}
where $P_2$ is the population of the excited state. 

To obtain the time development under $H_{\rm
c}$ of a wave packet incident from the left 
we solve first the stationary equation
\begin{equation}\label{eigenvalue}
H_{\rm  c}{\bf \Phi} = E {\bf \Phi},~~~~~{\rm where}~~{\bf
\Phi}(x)\equiv{\phi^{(1)}(x)\choose\phi^{(2)}(x)} 
\end{equation}
for scattering states 
with real energy 
$
E = \hbar^2k^2/2m \equiv E_k,  
$
which are incident from the left ($k>0$), 
\begin{equation}\label{A15}
{\bf \Phi}_k (x)= \frac{1}{\sqrt{2\pi}}
\left\{\begin{array}{ll}
\left(  
{e^{ikx}+ R_1e^{-ikx} \atop R_2 e^{-iqx}}
 \right), &\quad x\sim-\infty, 
\\ 
\left(  
{T_1 e^{ikx}\atop T_2 e^{iqx}}
 \right),& \quad x\sim\infty. 
\end{array}
\right. 
\end{equation}
These states are not orthogonal, in spite of the reality of $E$, 
because the Hamiltonian $H_{\rm c}$ 
is not Hermitian. The wavenumber $q$ obeys  
\begin{equation} \label{Eq}
E + i\hbar\gamma/2 = \hbar^2q^2/2m,  
\end{equation}
with ${\rm Im}\,q > 0$ for boundedness, while $R_{1,2}$ and $T_{1,2}$
are
reflection and transmission amplitudes 
for the ground and excited state  
channels.


If
$\widetilde{\psi}(k)$ denotes the wavenumber amplitude
that the wave packet
would have as a freely moving packet at $t=0$, then
\begin{equation}\label{2.9}
{\bf \Psi}(x,t) = \int_0^\infty dk \,\widetilde{\psi}(k) \,{\bf \Phi}_k
(x)\,e^{-i \hbar k^2 t/2m}
\end{equation}
describes the ``conditional''  
time development of the state for an undetected atom which in the
remote past comes in
from the left in the ground state.

Some simple forms of $\Omega(x)$  
admit analytical solutions for the stationary waves,
as demonstrated in 
\cite{DEHM02,DEHM03}, but we  
shall limit the present discussion
to an approximation that 
is valid for
``arbitrary'' shapes of $\Omega$  
within 
weak driving and low energy conditions \cite{ChY91}.     

Equation (\ref{eigenvalue}) reads explicitly 
\beqa
-\frac{\hbar^2}{2m}\frac{\partial^2\Phi^{(1)}(x)}{\partial x^2}
+\frac{\hbar}{2}
\Omega(x)\Phi^{(2)}(x)=E\Phi^{(1)}(x),
\label{1}
\\
-\frac{\hbar^2}{2m}\frac{\partial^2\Phi^{(2)}(x)}{\partial x^2}
+\frac{\hbar}{2}
\Omega(x)\Phi^{(1)}(x)-\frac{\hbar}{2}(2\Delta+i\gamma)
\Phi^{(2)}(x)=E\Phi^{(2)}(x). 
\label{2}
\eeqa
In the large $\gamma$ limit, 
\beq\label{condi}
\frac{\hbar|2\Delta +i\gamma|}{2}>>\frac{\hbar}{2}\Omega, E,
\eeq
we may neglect the kinetic and energy terms in the second equation  
to write 
\beq\label{21}
\Phi^{(2)}=\frac{\Omega}{2\Delta+i\gamma}\Phi^{(1)}.  
\eeq
Physically, 
it is assumed that the excited
state amplitude is small and proportional to the ground state 
amplitude, because the depletion by decay is rapid with respect to 
the atomic motion and to the Rabi pumping period. 
%
%
Substituting  (\ref{21}) into (\ref{1}), there results a
closed, one channel equation 
for $\Phi^{(1)}$ with 
a complex potential 
\beqa
V(x)&=&\frac{\hbar}{2}\frac{\Omega^2(x)}{2\Delta+i\gamma}
\\
&=&\frac{\hbar\Delta \Omega^2}{4\Delta^2+\gamma^2}
-i\,\frac{\hbar\gamma\Omega^2/2}{4\Delta^2+\gamma^2}, 
\eeqa
so that the sign of the real part has the sign of the detuning 
whereas the imaginary part is negative,
i.e. absorbing, for all $x$. 

The complex potential for the longitudinal direction $x$ 
and a traveling laser wave perpendicular to the initial 
atomic motion is similar to the one that appears 
for the transversal direction $y$ using the Raman-Nath
or related approximations to describe the incidence of the atom on
perpendicular standing laser fields \cite{ChY91,OABSZ96}.
In these approximations, however, $x$
is treated as 
a parameter according to $x=vt$ but $y$ is kept as a variable, whereas
here $x$ is the variable and we consider the motion along a ray of 
constant $y=0$. 

Notice that, within the stated conditions, the potential 
is independent of $E$. This implies that all (low energy) stationary 
scattering functions 
are subject to the same potential and therefore the time dependent 
Schr\"odinger equation also reduces to an effective
one channel equation with the effective potential $V$. 

%
%
%

Within the one channel approximation, we have to find scattering solutions
with the asymptotic form 
\begin{equation}
\Phi^{(1)}(x)= \frac{1}{\sqrt{2\pi}}
\left\{\begin{array}{ll}
{e^{ikx}+ R_1e^{-ikx}}
, &\quad x\sim-\infty, 
\\ 
{T_1 e^{ikx}}
,& \quad x\sim\infty. 
\end{array}
\right. 
\end{equation}
%
The absorption (i.e. detection) probability 
$A(k)$ for the incident $k$-plane wave
is given by 
\beq
A(k)=
1-|R_1|^2-|T_1|^2.   
\eeq

\section{Optimization}

The main aim of the present paper is to show that the atomic detection
may be improved by 
varying the spatial dependence of the laser intensity and 
the detuning.
To get optimal dependences for a fixed total length of the laser
illuminated
region
and for a given momentum range 
one has to find first, for weak driving conditions, the form 
of the complex potential that maximizes absorption.  
 
A very similar objective (optimizing a complex absorbing potential) 
is also pursued in time-dependent 
molecular scattering calculations to eliminate  
the outgoing wave packets at the edge of the computational box
and avoid unphysical effects \cite{MBM94,BMM94,PMS98,PM98a,PM98b}.  
We may take advantage of this coincidence by using, mutatis mutandis,
similar optimization techniques.     
An obvious difference with the molecular scattering 
case is that here the real and imaginary parts of the potential 
have specific physical meanings in terms of laser parameters.  
In particular the detuning $\Delta$ and Rabi frequency $\Omega$ 
are given by 
\beqa
\label{delta}
\Delta&=&-\frac{\gamma {\rm Re}(V)}{2{\rm Im}(V)},
\\
\Omega&=&|V|\left(\frac{2\gamma}{-\hbar {\rm Im}(V)}\right)^{1/2}.
\eeqa
This means that not all absorbing potentials $V$
can be admitted,
since the  
weak driving condition has to be satisfied, and moreover 
real and imaginary parts are not independent according
to Eq. (\ref{delta}).      
The limitation of the weak driving condition 
may be avoided, but at the price of loosing the 
simplified one-channel description, i.e., by keeping 
the two-channel equations and a $2\times2$
complex potential matrix. In the same manner one could also 
get rid of the 
limitation to small kinetic energies and optimize detection of
thermal atoms, for example. 
Here we shall discuss the simple one-channel case.

The strategy is variational: we choose a functional form 
for $V$ that depends on a few parameters and then find the
values of these parameters 
that maximize the absorption for a 
a given velocity interval, or, in practice, for a discrete set of 
$n$ velocities in such an interval \cite{MBM94},   
%
%
%
\beq\label{sum}
\bar{A}\equiv \sum_j A(k_j)W(k_j).    
\eeq
The ``weights'' $W(k_j)$ may be chosen  
according to the momentum 
distribution of the atoms, or, as in the examples discussed below, 
uniformly  for a chosen absorption 
window, $W(k_j)=1/n$    
In principle, it is possible to construct explicitly 
potentials that absorb perfectly 
at a discrete set of wavenumbers \cite{PMS98}, but these 
potentials tend to be too sensitive to small variations and 
thus ineffective for practical usage, and require in general 
arbitrary variations of real and imaginary parts. 
A further drawback in the present 
application is that they may have wild spatial variations or   
may correspond locally to strong driving conditions for which  
the one-channel, complex-potential model is not physically valid. 

Moreover, an ideal potential functional form for
optimizing atom detection should enable us  
to control the spatial scale in which the laser 
intensity 
varies significantly, in accordance with technical capabilities.
A simple form would be 
the sum of contiguous Gaussian functions with a certain width. 
Independently of the optimization algorithm used many evaluations of 
$\bar{A}$ are required in general. Each of them requires 
to solve numerically the Schr\"odinger equation to obtain the 
amplitudes $T_1$ and $R_1$, and 
this may be very time consuming.  
Here we shall choose an even simpler  
functional form, a set of contiguous ``square barriers'', as an 
approximation for a more realistic combination of Gaussians. 
At the present stage, where we are more interested in illustrating the 
concepts and general features involved than in any particular
application, this 
simplification does not introduce any significant distortion \cite{DEHM03}.       
The great numerical advantage of the square barriers is that 
the solution of the Schr\"odinger equation may be performed 
by multiplying 
a few $2\times 2$ transfer matrices, which is an extremely fast
process in comparison  
to the numerical techniques required for other functional forms.
Moreover one can write explicitly the gradient of $\bar{A}$ with respect 
to the real and imaginary parts of the barriers, so that very 
efficient optimization algorithms  
may be used.   
Expressions 
of the absorption and its gradient in terms of transfer matrices 
may be found in \cite{PMS98,PM98a,PM98b}. The main technical novelty 
here with respect to those works is the need to constrain
the real and imaginary parts according to Eq. (\ref{delta}) and to weak 
driving conditions. 
Our subroutine for constrained optimization 
is based in the successive quadratic programming algorithm
\cite{Schittkowski}.  

\section{Results and discussion}

\begin{figure}
\begin{center}
{\includegraphics[width=11cm]{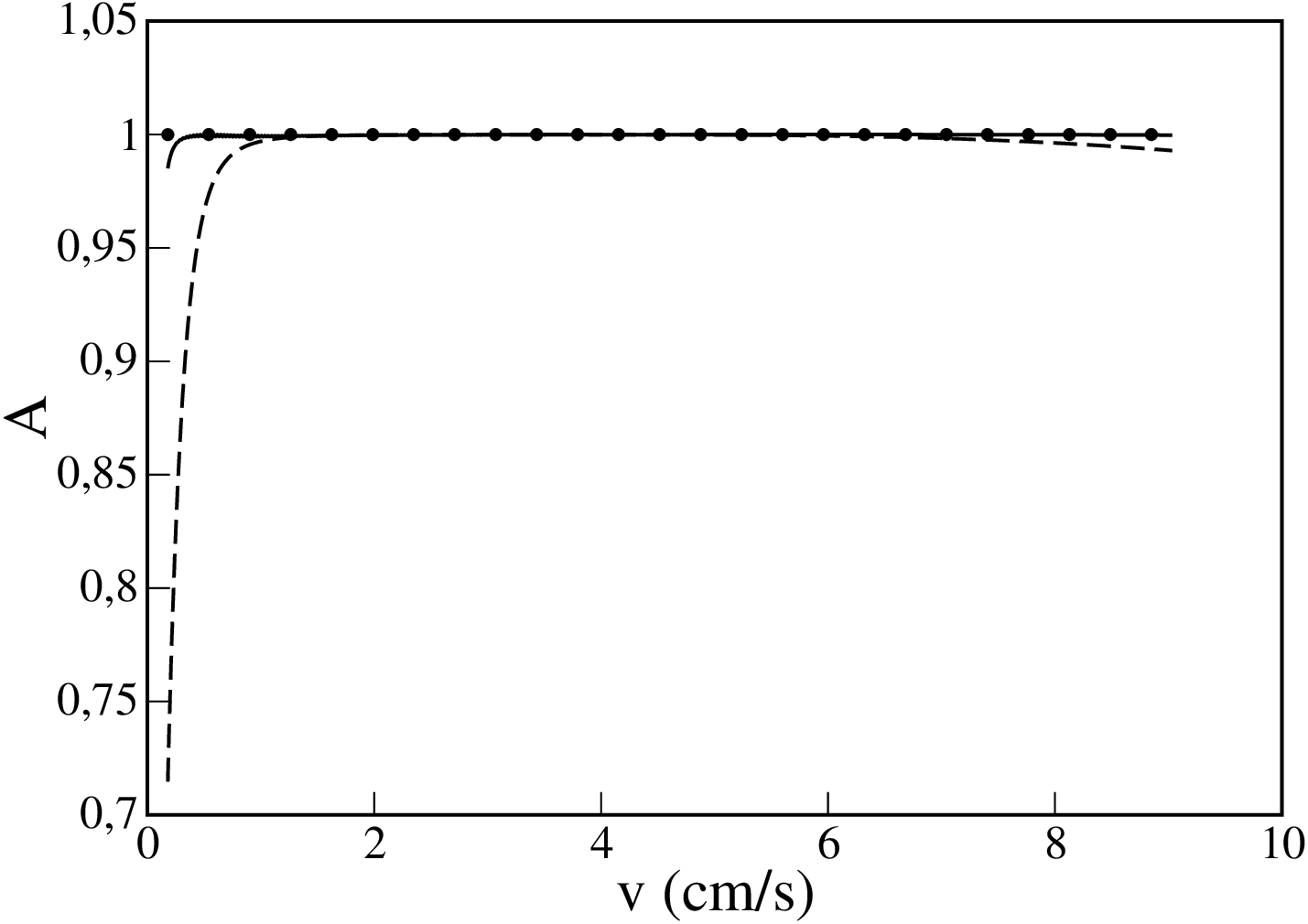}}
\caption{Absorption versus velocity $v$
for potentials optimized taking $n=100$ 
between $0.2$ and $9$ cm/s in (\ref{sum}). 
$L=10$ $\mu$m. Solid line: two barriers (each of 
5 $\mu$m); dashed line: one barrier; dots: eigth barriers.     
}
\label{f2}
\end{center}
\end{figure}

\begin{figure}
\begin{center}
{\includegraphics[width=11cm]{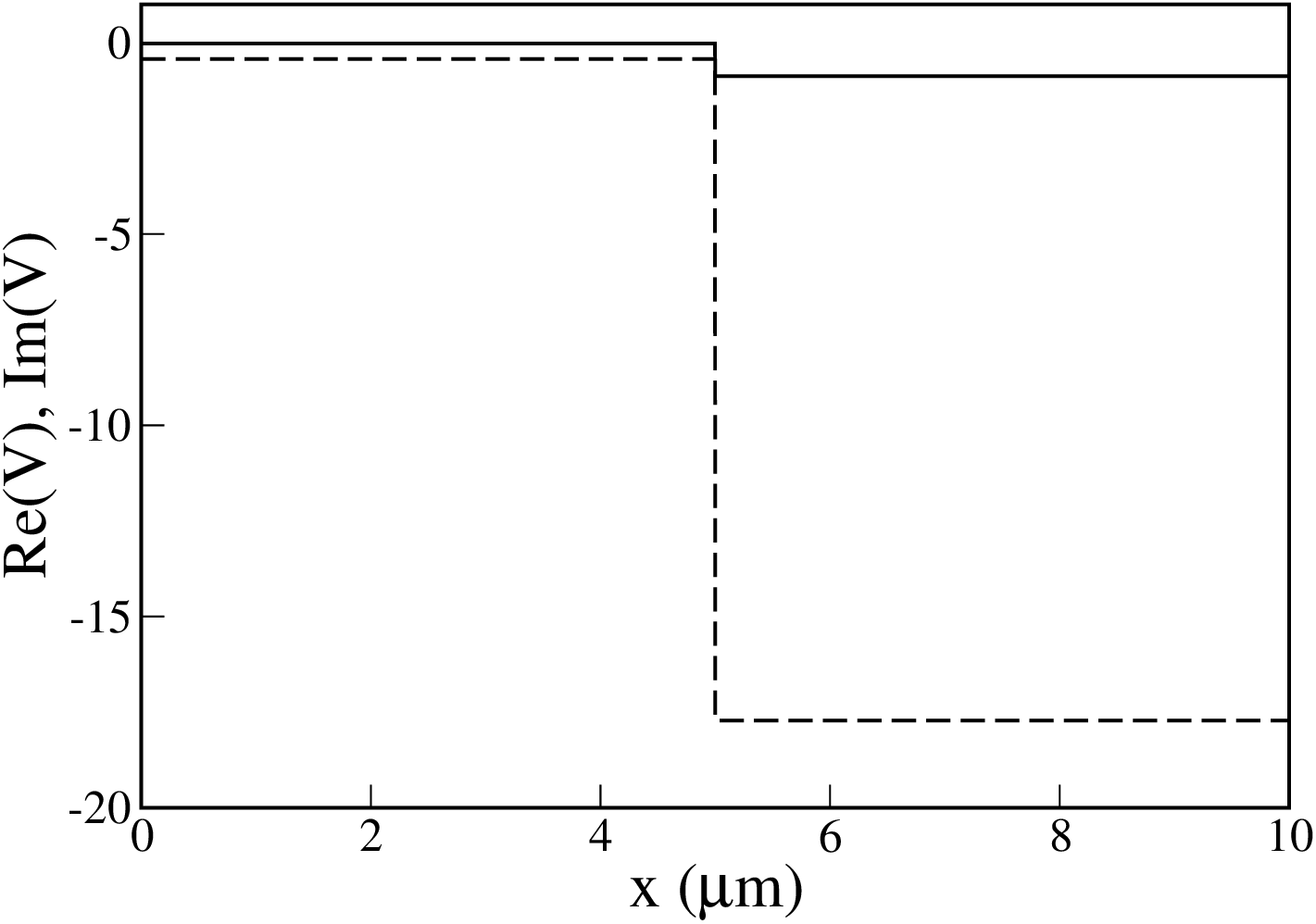}}
\caption{Real (solid line) and imaginary part (dashed line) 
of the two-barrier potential 
of Figure 2. 
}
\label{f3}
\end{center}
\end{figure}

\begin{figure}
\begin{center}
{\includegraphics[width=11cm]{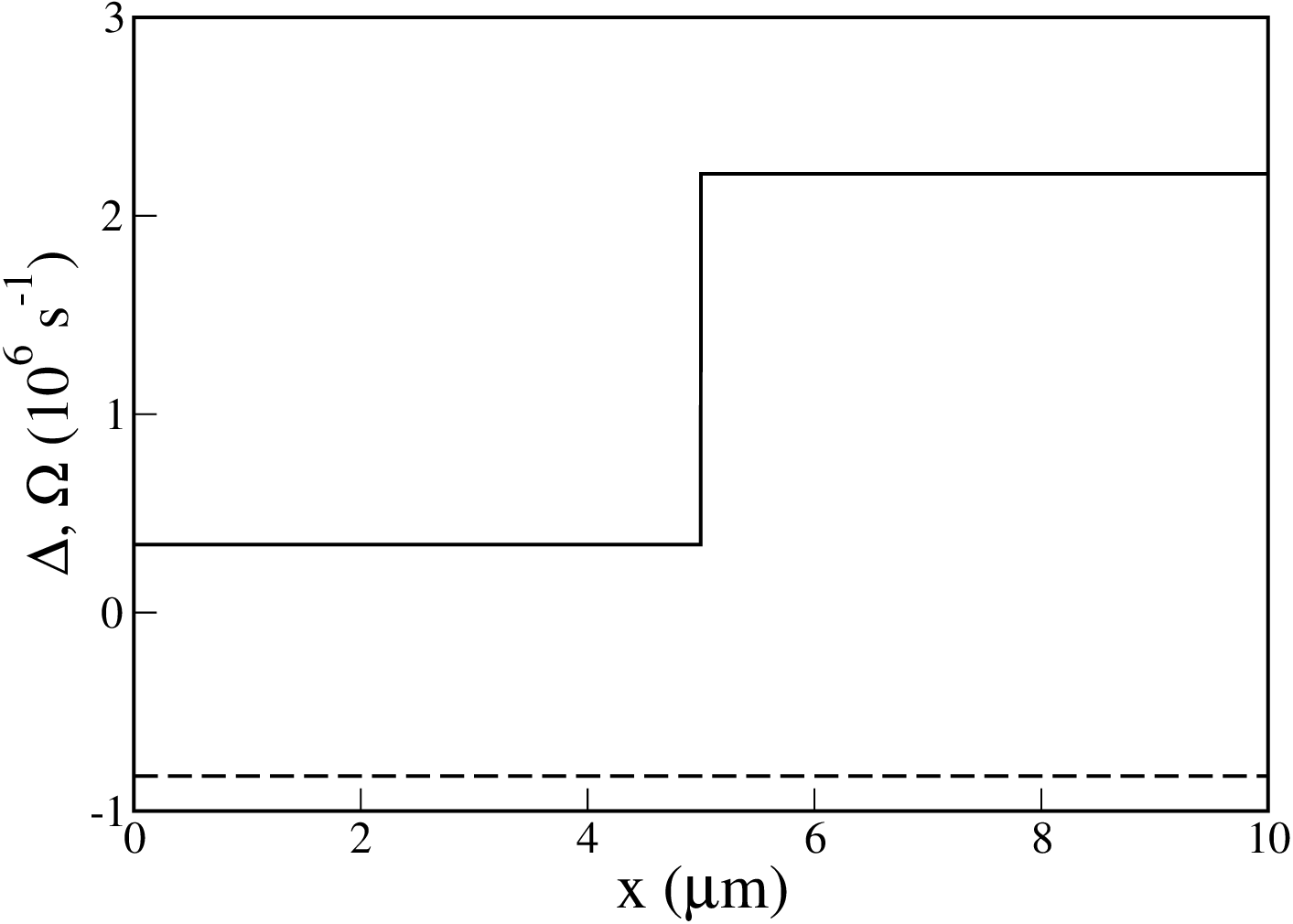}}
\caption{Detuning (dashed line) and Rabi frequency 
(solid line) of the two-barrier potential of
Figure 3.  
}
\label{f4}
\end{center}
\end{figure}

\begin{figure}
\begin{center}
{\includegraphics[width=11cm]{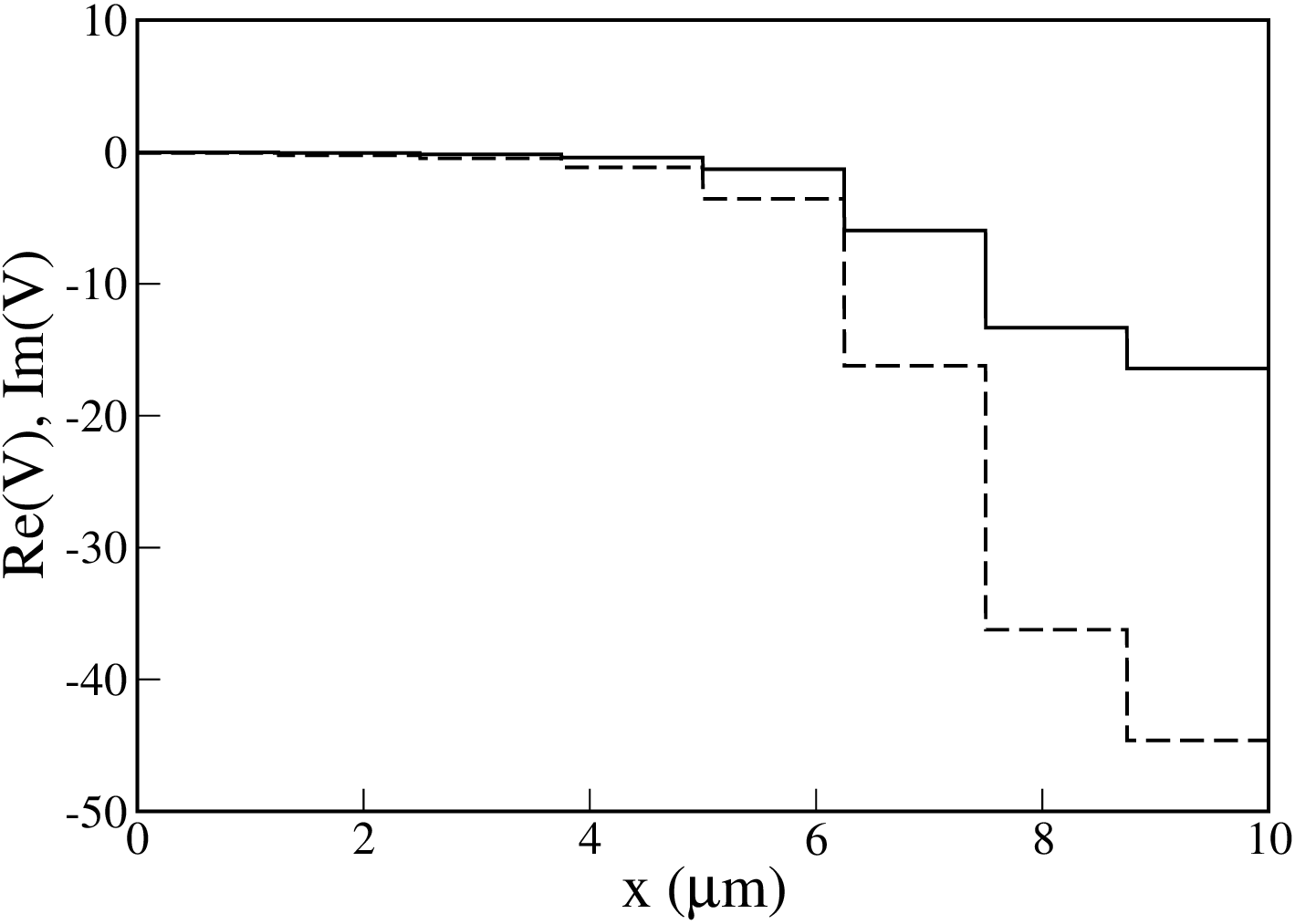}}
\caption{Real (solid line) and imaginary part (dashed line) 
of an optimized eight-barrier potential. Other parameters as in 
figure \ref{f2}.
}
\label{f5}
\end{center}
\end{figure}

\begin{figure}
\begin{center}
{\includegraphics[width=11cm]{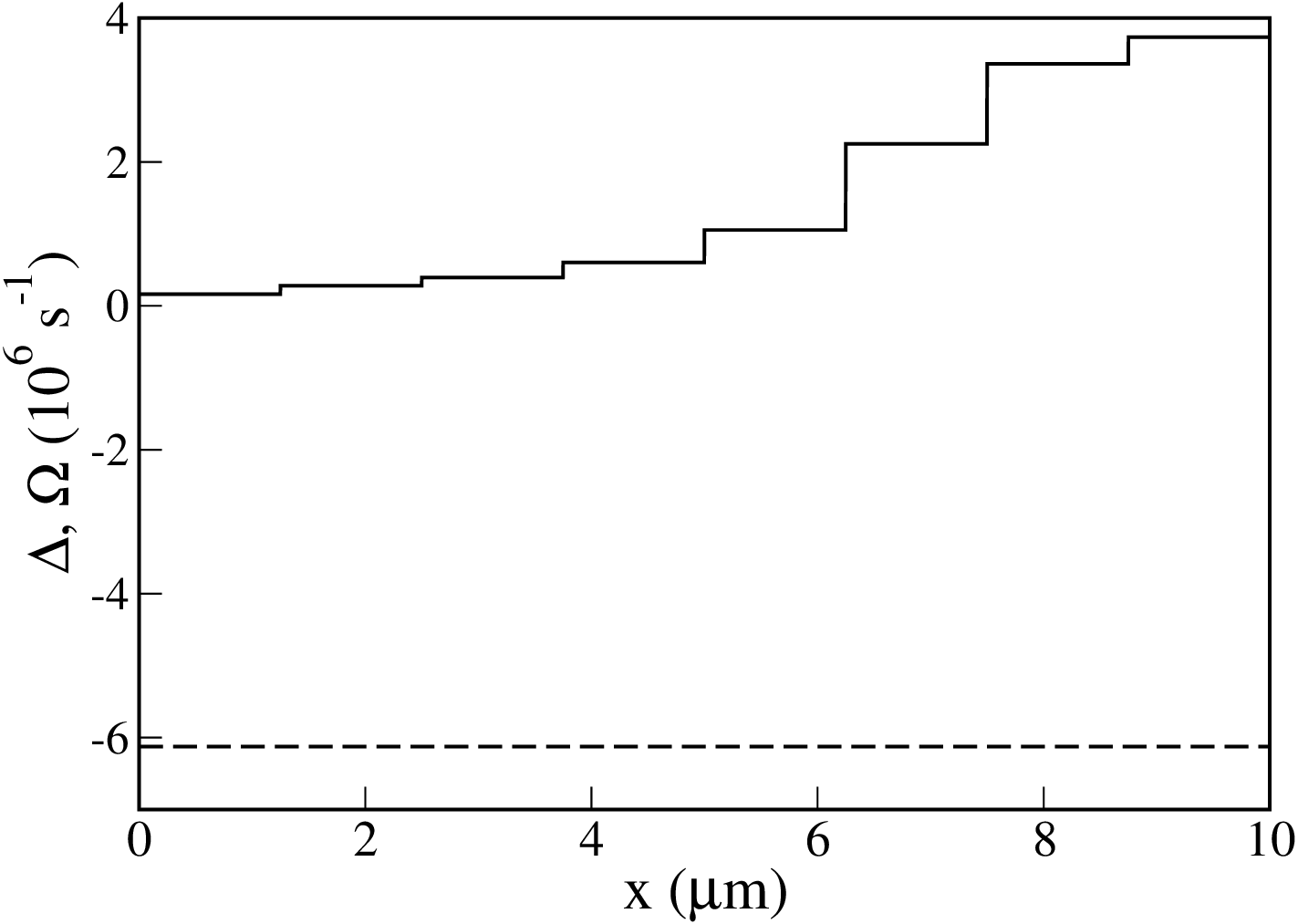}}
\caption{Detuning (dashed line) and Rabi frequency 
(solid line) of the eight-barrier potential of
figure \ref{f5}.  
}
\label{f6}
\end{center}
\end{figure}

Figure \ref{f2} shows the best 
absorption curves found for a range of momenta between
$0.2$ cm/s 
and $9$ cm/s considering one and two barriers for a total length 
of 10 $\mu$m of the laser illuminated region, as in Figure 1.
The potential, detuning and Rabi frequency 
corresponding to the optimal two-barrier case are shown in 
Figures \ref{f3} and \ref{f4}. 
Note that the optimal single barrier fails at both edges of the momentum 
range chosen whereas two barriers, with three free  parameters to 
optimize, do a much better job. 
We have also calculated an optimal eight barrier potential,
see figures \ref{f2}, \ref{f5}
and \ref{f6}, that suggests that the form of an ideal continuous potential 
would be characterized by a smooth, non-linear increase of laser intensity
with position, and negative detuning.

The present variational approach is based on a
number of simplifying assumptions but  
the concepts involved are applicable even outside the 
domain of validity of these assumptions. 
In particular, 
the restriction to weak laser interactions and low energies 
is not fundamental and may be removed by  
considering the full $2\times 2$ potential matrix instead of the 
one-channel effective potential. 
Similarly, more realistic and smooth functional forms 
may be used.   
We have assumed here that a number of 
laser beams with different intensity 
may be combined in a composite double (or multiple) beam
to maximize the detection probability,
but there are other simple 
possibilities to explore: Meneghini et al., for example, have considered 
a linear modulation of the detuning, that could be 
realized in a nonhomogeneous magnetic field, and a standard Gaussian 
form for the Rabi frequency \cite{MJLKSY00}.
This amounts to a four-parameter
functional 
form for the potential, whose absorption may be
also maximized for specific
applications.

Apart from maximizing detection, there are other quantities
that could be 
maximized or minimized, such as the detection delay. 
A minimal detection delay would be of interest for an accurate 
measurement of arrival times \cite{DEHM02,DEHM03,ML00}. 
One possible application of a ``perfect detector'' in a broad
momentum range,
would be the 
measurement, for the first time, of the backflow effect, 
namely, negative current densities 
for an atomic wave packet 
composed by positive momenta \cite{DEHM02,ML00,backflow,Allcock69,MPL99}.   
This and other extensions of the present work will be dealt with 
elsewhere.

\ack{We are grateful to J. P. Palao and S. Brouard for technical 
assistance.  
This work has been supported
by Ministerio de Ciencia y Tecnolog\'\i a (BFM2000-0816-C03-03), 
UPV-EHU (00039.310-13507/2001), and a German-Spanish 
collaboration Grant.  
}


\end{document}